\documentclass[a4paper,11pt]{article}%
\usepackage{amssymb,amsmath,amsfonts}
\usepackage[dvips]{graphicx}
\usepackage{amsmath}
\usepackage{amsfonts}
\usepackage{amssymb}%
\providecommand{\U}[1]{\protect\rule{.1in}{.1in}}
\providecommand{\U}[1]{\protect\rule{.1in}{.1in}}
\newtheorem{theorem}{Theorem}

\newtheorem{comment}[theorem]{Comment}

\newtheorem{idea memo}[theorem]{Idea Memo}

\begin{document}

\title{A Unified Scheme of \\Measurement and Amplification Processes \\based on Micro-Macro Duality \\-- Stern-Gerlach experiment as a typical example --}
\author{Ryo HARADA and Izumi OJIMA\\Research Institute for Mathematical Sciences, \\Kyoto University, Kyoto 606-8502, Japan}
\maketitle

\begin{abstract}
A unified scheme for quantum measurement processes is formulated on the basis
of Micro-Macro duality as a mathematical expression of the general idea of
\textit{quantum-classical correspondence}. In this formulation, we can
naturally accommodate the amplification processes necessary for magnifying
quantum state changes at the microscopic end of the probe system into the
macroscopically visible motion of the measuring pointer. Its essence is
exemplified and examined in the concrete model of the Stern-Gerlach experiment
for spin measurement, where the Helgason duality controlling the Radon
transform is seen to play essential roles.

\end{abstract}

\section{Introduction: Micro-Macro Duality and Measurements}

\label{Intro}

In this paper, we present a unified description of a measurement process of
quantum observables together with the amplification process associated with
it. For this purpose, we recall the essence of Micro-Macro duality
\cite{Unif03, Ojima, O-T} as a mathematical expression of the general idea of
quantum-classical correspondence which plays crucial roles. In this context,
we note that the `boundary' between the quantum and classical levels can be
found in the notion of a sector, in terms of which we can understand, in a
clear-cut manner, the mutual relations between the microscopic quantum world
and the macroscopic classical levels. To define a sector, we classify
representations and states of a C*-algebra $\mathfrak{A}$ of quantum
observables according to the \textit{quasi-equivalence} $\pi_{1}\approx\pi
_{2}$ \cite{Dix} defined by the unitary equivalence of representations
$\pi_{1},\pi_{2}$ \textit{up to multiplicity}, which is equivalent to the
isomorphism of von Neumann algebras $\pi_{1}(\mathfrak{A}\mathfrak{)}%
^{\prime\prime}\simeq\pi_{2}(\mathfrak{A}\mathfrak{)}^{\prime\prime}$ of
representatoins $\pi_{1}$ and $\pi_{2}$. A \textit{sector} or a \textit{pure
phase} in the physical context is then defined by a quasi-equivalence class of
\textit{factor} representations and states corresponding to a von Neumann
algebra with a trivial centre, which is a minimal unit among quasi-equivalence
classes. Representations belonging to different sectors $\pi_{a}$ and $\pi
_{b}$ are mutually \textit{disjoint }with no non-zero intertwiners: namely, if
$T$ is an intertwiner from $\pi_{a}$ to $\pi_{b}$ defined as a bounded
operator $T$ from the representation space $\mathfrak{H}_{\pi_{a}}$ of
$\pi_{a}$ to that $\mathfrak{H}_{\pi_{b}}$ of $\pi_{b}$ satisfying the
relation $T\pi_{a}(A)=\pi_{b}(A)T$ ($\forall A\in\mathfrak{A}$), then it
vanishes, $T=0$.

If $\pi$ is not a factor representation belonging to one sector, it is called
a \textit{mixed phase.} In the standard situations where separable Hilbert
spaces are used, a mixed phase can uniquely be decomposed into a direct sum
(or integral) of sectors, through the spectral decomposition of its
non-trivial centre $\mathfrak{Z}(\pi(\mathfrak{A}\mathfrak{)}^{\prime\prime
})=\pi(\mathfrak{A})^{\prime\prime}\cap\pi(\mathfrak{A})^{\prime
}=:\mathfrak{Z}_{\pi}(\mathfrak{A})$ of $\pi(\mathfrak{A}\mathfrak{)}%
^{\prime\prime}$ which is a commutative von Neumann algebra admitting a
`simultaneous diagonalization'. Each sector contained in $\pi$ is faithfully
parametrized by the Gel'fand spectrum $Spec(\mathfrak{Z}_{\pi}(\mathfrak{A}))$
of the centre $\mathfrak{Z}_{\pi}(\mathfrak{A})$. Thus, commutative classical
observables belonging to the centre physically play the role of
\textit{macroscopic order parameters} and the central spectrum
$Spec(\mathfrak{Z}_{\pi}(\mathfrak{A}))$ can be regarded as the
\textit{classifying space of sectors} to register faithfully all the sectors
contained in $\pi$. In this way, we find in a \textit{mixed phase} $\pi$ the
coexistence of quantum (=\textit{intra-sectorial}) and classical systems, the
latter of which describes an \textit{inter-sectorial} structure in terms of
order parameters constituting the centre $\mathfrak{Z}_{\pi}(\mathfrak{A})$.
In this way, the `boundary' and the gap between the quantum world described by
non-commutative algebras of quantum variables and the classical levels with
commutative algebras of order parameters can be identified with a
(\textit{superselection})\textit{ sector structure }consisting of a family of
sectors or pure phases \cite{Unif03}.

\subsection*{Intra-sectorial Analysis by MASA}

Since a single sector or a pure phase corresponds to a (quasi-equivalence
class of) factor representation $\pi$ of a C*-algebra $\mathfrak{A}$ of
quantum observables, its \textit{intra-sectorial} structure, the structure
inside of a sector, is described by the observables belonging to the factor
von Neumann algebra $\mathcal{M}=\pi(\mathfrak{A})^{\prime\prime}$
corresponding to $\pi$. In this and the next sections, we recapitulate the
essence of the general scheme to analyze the intra-sectorial structure
\cite{Ojima, O-T}. Because of the non-commutativity of $\mathcal{M}$, what can
be experimentally observed through a measurement is up to a certain maximal
abelian subalgebra (MASA, for short) $\mathcal{A}=\mathcal{A}^{\prime}%
\cap\mathcal{M}$ (with $\mathcal{A}^{\prime}$ the commutant of $\mathcal{A}$)
of $\mathcal{M}$: elements of a MASA $\mathcal{A}$ can be regarded as
macroscopic observables to visualize some aspects of the microscopic structure
of a sector in the macroscopic form of $Spec(\mathcal{A})$. In fact, a tensor
product $\mathcal{M}\otimes\mathcal{A}$ (acting on the tensor product Hilbert
space $\mathfrak{H}_{\omega}\otimes L^{2}(Spec(\mathcal{A}))$) has a centre
given by $\mathfrak{Z}(\mathcal{M}\otimes\mathcal{A})=\mathfrak{Z}%
(\mathcal{M})\otimes\mathcal{A}=\mathbf{1}\otimes L^{\infty}(Spec(\mathcal{A}%
))$, and hence, the spectrum $Spec(\mathcal{A})$ of a MASA $\mathcal{A}$ to be
measured can be understood as parametrizing a \textit{conditional sector
structure} of the composite system $\mathcal{M}\otimes\mathcal{A}$ of the
observed system $\mathcal{M}$ and $\mathcal{A}$, the latter of which can be
identified with the measuring apparatus $\mathcal{A}$ in the simplified
version \cite{Unif03} of Ozawa's measurement scheme \cite{Oza}. This picture
of conditional sector structure is consistent with the physical essence of a
measurement process as `classicalization' of some restricted aspects
$\mathcal{A}$($\subset\mathcal{M}$) of a quantum system, conditional on the
coupling $\mathcal{M}\otimes\mathcal{A}$ of $\mathcal{M}$ with the apparatus
identified with $\mathcal{A}$.

To implement a physical process to measure the observables in $\mathcal{A}$,
we need to specify a dynamical coupling between the observed and measuring
systems, which is accomplished by choosing such a unitary group $\mathcal{U}$
in $\mathcal{A}$ as generating $\mathcal{A}$, i.e., $\mathcal{A}%
=\mathcal{U}^{\prime\prime}$. In the standard situation where the relevant
Hilbert space is separable, the abelian von Neumann algebra $\mathcal{A}$ on
it is generated by a single element, and hence, we can assume without loss of
generality that $\mathcal{U}$ is a locally compact abelian Lie group. Because
of the commutativity of $\mathcal{U}$, the group characters $\gamma$ of
$\mathcal{U}$, $\gamma:\mathcal{U}\ni u\longmapsto\gamma(u)\in\mathbb{T}$ (:
1-dimensional torus) s.t. $\gamma(u_{1}u_{2})=\gamma(u_{1})\gamma(u_{2})$,
$\gamma(e)=1$, constitute the dual goup $\widehat{\mathcal{U}}$ satisfying the
Fourier-Pontryagin duality $\hat{\hat{\mathcal{U}}}\simeq\mathcal{U}$. Since
the restriction $\chi\upharpoonright_{\mathcal{U}}$ to $\mathcal{U\subset A}$
of an \textit{algebraic character} $\chi\in Spec(\mathcal{A})$ of
$\mathcal{A}$ is naturally a \textit{group character} of $\mathcal{U}$, a
canonical embedding $Spec(\mathcal{A})\hookrightarrow\widehat{\mathcal{U}}$
can be defined by $Spec(\mathcal{A})\ni\chi\longmapsto\chi\upharpoonright
_{\mathcal{U}}\in\widehat{\mathcal{U}}$. As the MASA $\mathcal{A}%
=\mathcal{A}^{\prime}\cap\mathcal{M}$ is the fixed-point subalgebra
$\mathcal{A}=\mathcal{M}^{Ad(\mathcal{U)}}$ of $\mathcal{M}$ under the adjoint
action of $\mathcal{U}$, our discussion can also be related with the
Galois-theoretical context of the duality between W*-dynamical systems
$\mathcal{M}\underset{\alpha}{\curvearrowleft}\mathcal{U}$ and $\mathcal{M}%
^{\alpha(\mathcal{U})}\underset{\hat{\alpha}}{\curvearrowleft}\widehat
{\mathcal{U}}$ and between the associated crossed products $\mathcal{M}%
\rtimes_{\alpha}\mathcal{U}$ and $\mathcal{M}^{\alpha(\mathcal{U})}%
\rtimes_{\hat{\alpha}}\widehat{\mathcal{U}}$, where the co-action
$\widehat{\alpha}$ of $\mathcal{U}$ dual to $\alpha$ can be identified with an
action of $\widehat{\mathcal{U}}$: $\mathcal{U}\underset{\alpha}%
{\curvearrowright}\mathcal{M(}\simeq\mathcal{M}^{\alpha(\mathcal{U})}%
\rtimes_{\hat{\alpha}}\widehat{\mathcal{U}})\rightleftarrows(\mathcal{M}%
\rtimes_{\alpha}\mathcal{U)}\underset{\hat{\alpha}}{\curvearrowleft}%
\widehat{\mathcal{U}}$. This co-action $\hat{\alpha}$ plays important roles in
the reconstruction of quantum (microscopic) systems from the classical
macroscopic data.

\section{Measurement Coupling, Kac-Takesaki Operators and Instrument}

\label{Measurement}

We show that the above measurement coupling can be specified by means of a
Kac-Takesaki operator \cite{Tats} (K-T operator, for short), one of the
central notions in harmonic analysis (where it is called a fundamental
operator in \cite{Enock} and a multiplicative unitary in \cite{BaajSkan}). In
what follows this operator is seen to play essential roles in our whole scheme
to unify both measurement and amplification processes. In the regular
representation of the group $\mathcal{U}$, a K-T operator $W$ is defined by
\[
(W\eta)(u,v):=\eta(v^{-1}u,v)
\]
for $\eta\in L^{2}(\mathcal{U}\times\mathcal{U},du\otimes du),u,v\in
\mathcal{U}$ with $du$ the Haar measure of $\mathcal{U}$, characterized by the
pentagonal and intertwining relations:
\begin{align*}
W_{12}W_{23}  &  =W_{23}W_{13}W_{12},\\
W(1\otimes\lambda_{u})  &  =(\lambda_{u}\otimes\lambda_{u})W,
\end{align*}
where the suffices $1,2,3$ indicate the places in the tensor product
$L^{2}(\mathcal{U},du)\otimes L^{2}(\mathcal{U},du)\otimes L^{2}%
(\mathcal{U},du)$ on which the operators act.

The simplest form of the action $\alpha$, $\mathcal{M}\underset{\alpha
}{\curvearrowleft}\mathcal{U}$, of $\mathcal{U}$ on $\mathcal{M}$ is given by
the adjoint action $\alpha_{u}(M)=Ad_{u}(M)=uMu^{-1}$, as commonly found in
many discussions on the measurement processes. This corresponds physically to
such an approximation to the coupled dynamics of the composite system
$\mathcal{M}\otimes\mathcal{A}$ that the Hamiltonian $H_{0}$ intrinsic to the
observed system is neglected but the bilinear coupling $H_{I}=\lambda\sum
_{i}X_{i}\otimes A_{i}$ is kept between the system observables $X_{i}$
($\exp(iX_{i})\in\mathcal{M}$) and the external forces $A_{i}$ ($\exp
(iA_{i})\in\mathcal{A}$). To retain the effects of the dynamics intrinsic to
the observed system, we take here a more general form of the action
$\mathcal{M}\underset{\alpha}{\curvearrowleft}\mathcal{U}$ of the measuring
system than the adjoint one under the assumption that $\alpha$ is unitarily
implemented, $\alpha_{u}(M)=U_{u}MU_{u}^{-1}$ ($M\in\mathcal{M}$,
$u\in\mathcal{U}$), by a unitary representation $U$ of $\mathcal{U}$ on the
standard representation Hilbert space $L^{2}(\mathcal{M})$ of $\mathcal{M}$.
Then the representation $U(W)$ of $W$ corresponding to $\alpha=Ad(U)$ is
defined by%
\[
(U(W)\xi)(u):=U_{u}(\xi(u))\text{ \ \ \ for }\xi\in L^{2}(\mathcal{M})\otimes
L^{2}(\mathcal{U},du),
\]
satisfying the pentagonal and intertwining relations:%
\begin{align*}
U(W)_{12}W_{23}  &  =W_{23}U(W)_{13}U(W)_{12},\\
U(W)(1\otimes\lambda_{u})  &  =(U_{u}\otimes\lambda_{u})U(W).
\end{align*}
The meaning of $U(W)$ can be seen in the following heuristic expression in
Dirac's bra-ket notation:
\[
U(W)=\int_{u\in\mathcal{U}}U_{u}\otimes|u\rangle du\langle u|.
\]

This unitary operator $U(W)$ provides the coupling between the observed and
measuring systems precisely required for measuring the observables in
$\mathcal{A}$. For this purpose, we examine the action of its Fourier
transform on the state vectors of the composite system belonging to
$L^{2}(\mathcal{M})\otimes L^{2}(\mathcal{U},du)$. First, in terms of the
Fourier transform $(\mathcal{F}\xi)(\gamma):=\int_{\mathcal{U}}\overline
{\gamma(u)}\xi(u)du$ for $\xi\in L^{2}(\mathcal{U},du)$, the Fourier transform
$V:=(\mathcal{F}\otimes\mathcal{F})W^{\ast}(\mathcal{F}\otimes\mathcal{F}%
)^{-1}$ of the K-T operator $W$ on $L^{2}(\mathcal{U}\times\mathcal{U})$ is
defined, which turns out just to be the K-T operator of the dual group
$\widehat{\mathcal{U}}$ (equipped with the Plancherel measure $d\hat{u}$)
satisfying and characterized by the relations:
\begin{align*}
(V\eta)(\gamma,\chi)  &  =\eta(\gamma,\gamma^{-1}\chi)\text{ \ \ \ for }%
\eta\in L^{2}(\widehat{\mathcal{U}},d\hat{u})\text{,}\\
V_{23}V_{12}  &  =V_{12}V_{13}V_{23},\\
V(\lambda_{\gamma}\otimes1)  &  =(\lambda_{\gamma}\otimes\lambda_{\gamma})V.
\end{align*}
Similarly, the Fourier transform of $\mathcal{U}(W)$ is defined by
$\widetilde{U}(V):=(id\otimes\mathcal{F})U(W)^{\ast}(id\otimes\mathcal{F}%
)^{-1}$. Owing to the SNAG theorem due to the abelianness of $\mathcal{U}$,
its unitary representation $\mathcal{U}\ni u\longmapsto U_{u}\in
\mathcal{U}(L^{2}(\mathcal{M}))$ admits the spectral decomposition $U_{u}%
=\int_{\chi\in Spec(\mathcal{A})\subset\mathcal{\hat{U}}}\overline{\chi
(u)}dE(\chi)$, corresponding to which $\widetilde{U}(V)$ has the spectral
decompostion given by%
\[
\widetilde{U}(V)=\int_{\chi\in Spec(\mathcal{A})}dE(\chi)\otimes\lambda_{\chi
}.
\]
In the Dirac notation, the action of $\widetilde{U}(V)$ on $L^{2}%
(\mathcal{M})\otimes L^{2}(\widehat{\mathcal{U}})$ is given for $\gamma
\in\widehat{\mathcal{U}}$, $\xi\in L^{2}(\mathcal{M})$, by%
\begin{equation}
\widetilde{U}(V)(\xi\otimes|\gamma\rangle)=\int_{\chi\in Spec(\mathcal{A}%
)}dE(\chi)\xi\otimes|\chi\gamma\rangle. \label{eqn:FT}%
\end{equation}
To understand the physical meaning of the above quantities, we introduce some
such vocabularies \cite{Oza} as `probe system' and `neutral position' in
measurement processes: the former means the \textit{microscopic end} of the
measuring apparatus \textit{at its microscopic contact point} with the
observed system, and the latter the \textit{initial} (\textit{microscopic})
\textit{state of the probe system corresponding to the macroscopically stable
position of the measuring pointer realized when the apparatus is isolated}.

To see clearly the essence of the formulation, we assume that $\widehat
{\mathcal{U}}$ is discrete (or, equivalently, $\mathcal{U}$ is compact); then
we can plug into $\gamma\in Spec(\mathcal{A})(\subset\widehat{\mathcal{U}})$
and $\xi\in L^{2}(\mathcal{M})$ in Eq.(\ref{eqn:FT}), respectively, the group
identity $\iota\in\widehat{\mathcal{U}}$ and such an eigenstate $\xi=\xi
_{\chi}$ as $A\xi_{\chi}=\chi(A)\xi_{\chi}$ ($\forall A\in\mathcal{A}$) of
$\chi\in Spec(\mathcal{A})$, which gives%
\begin{equation}
\widetilde{U}(V)(\xi_{\chi}\otimes|\iota\rangle)=\xi_{\chi}\otimes|\chi
\rangle. \label{eqn:FT1}%
\end{equation}
Namely, corresponding to the eigenstate $\xi_{\chi}$ of $\mathcal{A}$ found in
the observed system, the coupling unitary $\widetilde{U}(V)$ causes such a
state change as $|\iota\rangle\rightarrow|\chi\rangle$ in the probe system.
For such a generic state as $\xi=\sum_{\chi\in Spec(\mathcal{A})}c_{\chi}%
\xi_{\chi}$ of the observed system, therefore, we obtain%
\begin{equation}
\widetilde{U}(V)(\xi\otimes|\iota\rangle)=\sum_{\chi\in\widehat{\mathcal{U}}%
}c_{\chi}\xi_{\chi}\otimes|\chi\rangle, \label{eqn:FT2}%
\end{equation}
that is, the unitary operator $\widetilde{U}(V)$ creates from a decoupled
state $\xi\otimes|\iota\rangle$ of $\mathcal{M}\otimes\mathcal{A}$ a `perfect
correlation' \cite{Oza05} between states of the observed system and of the
probe system, which is just required for transmitting the information from the
observed system to the probe system. When the group $\mathcal{U}$ is not
compact with $\widehat{\mathcal{U}}$ not being discrete, the identity element
$\iota\in\widehat{\mathcal{U}}$ is not represented by a normalized vector,
$|\iota\rangle\notin L^{2}(\mathcal{U})$, but we can choose an invariant mean
$m_{\mathcal{U}}$ over $\mathcal{U}$ owing to the amenability of the abelian
group $\mathcal{U}$ which plays the physically equivalent roles of the neutral
position $\iota$. As all what can be realized in this case is known \cite{Oza}
to be the \textit{approximate measurements}, the formula corresponding to
Eq.(\ref{eqn:FT2}) can be given by Eq. (\ref{eqn:FT}) and by the use of
$m_{\mathcal{U}}$ as seen below in Eq. (\ref{instru}). In this way the K-T
operators are seen to fullfil the necessary tasks for materializing the
physical essence of measurements in the mathematical formulation: the K-T
operator $U(W)$ determines the coupling between the observed and the measuring
systems and its Fourier transform $\widetilde{U}(V)$ given by Eq.(\ref{eqn:FT}%
) establishes the `perfect correlation' \cite{Oza05}.

Integrating all the ingredients relevant to our measurement scheme, we define
an instrument $\mathcal{I}$ as a completely positive operation-valued measure
as follows:
\begin{align}
\mathcal{I}(\Delta|\omega_{\xi})(B)  &  :=(\omega_{\xi}\otimes m_{\mathcal{U}%
})\left(  \widetilde{U}(V)^{\ast}(B\otimes\chi_{\Delta})\widetilde
{U}(V)\right) \nonumber\\
&  =(\langle\xi|\otimes\langle\iota|)\widetilde{U}(V)^{\ast}(B\otimes
\chi_{\Delta})\widetilde{U}(V)(|\xi\rangle\otimes|\iota\rangle)\nonumber\\
&  =\int_{\Delta}\sqrt{\frac{dE(\gamma)}{d\mu(\gamma)}}B\sqrt{\frac
{dE(\gamma)}{d\mu(\gamma)}}d\mu(\gamma)=:\int_{\Delta}\sqrt{dE(\gamma)}%
B\sqrt{dE(\gamma)}, \label{instru}%
\end{align}
where $\omega_{\xi}$ s.t. $\omega_{\xi}(B)=\langle\xi|B\xi\rangle$ is an
initial state of the observed system, $d\mu(\gamma)$ an arbitrary probability
measure with respect to which the spectral measure $dE(\gamma)$ of $U$ is
absolutely continuous: $dE(\gamma)\ll d\mu(\gamma)$, and $\chi_{\Delta}$ the
indicator function of a Borel set $\Delta\subset Spec(\mathcal{A})$ to which
the measured values of $\mathcal{A}$ belongs. The spectral measure
$dE(\gamma)$ is just the \textit{effect} of the measurement, from which our
K-T operator $\widetilde{U}(V)$ can be reconstructed by $\widetilde{U}%
(V)=\int_{\chi\in Spec(\mathcal{A})}dE(\chi)\otimes\lambda_{\chi}$. In this
sense, the \textit{three notions}, the K-T operator $\widetilde{U}(V)$, the
effect $dE(\gamma)$ and the instrument $\mathcal{I}(\Delta|\omega_{\xi})$,
\textit{are all mutually equivalent}. The most important essence of the
statistical interpretation in the measurement processes is summarized in this
notion of instrument as follows: the probability distribution for measured
values of observables in $\mathcal{A}$ to be found in a Borel set
$\Delta\subset Spec(\mathcal{A})$ is given by $p(\Delta|\omega)=\mathcal{I}%
(\Delta|\omega)(\mathbf{1})$ and, associated with this, the initial state
$\omega$ of the observed system is changed by the read-out of measured values
in $\Delta$ into a final state given in such a form as $\mathcal{I}%
(\Delta|\omega_{\xi})/p(\Delta|\omega_{\xi})$ \cite{Oza}, according to which a
process of the so-called `reduction of wave packets' is described.
Incidentally, the reason for the relevance of the \textit{Fourier transform}
from $\mathcal{U}(W)$ to $\widetilde{U}(V)=(id\otimes\mathcal{F})U(W)^{\ast
}(id\otimes\mathcal{F})^{-1}$ can naturally be understood in relation with the
duality between the (algebra of) observables and the states: when the group
$\mathcal{U}$ acts on the algebra $\mathcal{M}$ of the observed system, the
corresponding states can be parametrized by~$\widehat{\mathcal{U}}$ as
\textquotedblleft eigenstates\textquotedblright\ w.r.t. the action $U$ of
$\mathcal{U}$, which should also be read out as the measured values.

By means of the instrument $\mathcal{I}$, a measurement process is described
as the process of state changes due to the measurement coupling $\alpha=Ad(U)$
which transforms an \textit{initial} state $\omega$ of the observed system
\textit{decoupled }from the probe system into \textit{final} ones of the same
nature, in parallel with the \textit{scattering processes} described in terms
of the \textit{incoming} and \textit{outgoing} asymptotic states of free
particles. The algebra describing the composite system is the tensor algebra
$\mathcal{M}\otimes\mathcal{A}=\mathcal{M}\otimes L^{\infty}(Spec(\mathcal{A}%
))$ realized in the \textit{initial} and \textit{final} stages, respectively,
before and after the measuring processes according to the \textit{switching-on
}and\textit{ -off of the coupling} $\alpha=Ad(U)$. As incoming and outgoing
asymptotic fields, $\varphi^{in}$ and $\varphi^{out}$, in quantum field theory
are interpolated by \textit{interacting Heisenberg fields }$\varphi_{H}$, we
can consider a similar description of the composite system of $\mathcal{M}$
and $\mathcal{A}$ with the coupled dynamics $\alpha$ incorporated at the level
of the algebra which interpolates the initial and final decoupled system
$\mathcal{M}\otimes\mathcal{A}$. This is given by the notion of the crossed
products $\mathcal{M}\rtimes_{\alpha}\mathcal{U}$ of the algebra due to the
action $\alpha$ of $\mathcal{U}$ on $\mathcal{M}$, in terms of which the
effect of the measuring coupling in the measurement process can be seen in
such a form as $\mathcal{M}\otimes\mathcal{A}=\mathcal{M}\rtimes
_{\alpha=id_{\mathcal{M}}}\mathcal{U}\rightarrow\mathcal{M}\rtimes_{\alpha
}\mathcal{U}\rightarrow\mathcal{M}\otimes\mathcal{A}$, in parallel with the
scattering processes, $\varphi^{in}\rightarrow\varphi_{H}\rightarrow
\varphi^{out}$. In terms of the K-T operators, the crossed product
$\mathcal{M\rtimes_{\alpha}U}$ as an important notion in the Fourier-Galois
duality is defined on $L^{2}(\mathcal{M})\otimes L^{2}(\mathcal{U})$ in the
following two equivalent ways: either as a von Neumann algebra $\lambda
^{\mathcal{M}}(L^{1}(\mathcal{U},\mathcal{M}))^{\prime\prime}$ generated by
the Fourier transform $\lambda^{\mathcal{M}}(\hat{F}):=\int_{\mathcal{U}}%
\hat{F}(u)U(u)du$ of $\mathcal{M}$-valued $L^{1}$-functions $\hat{F}\in
L^{1}(\mathcal{U},\mathcal{M})$ with the convolution product, $(\hat{F}%
_{1}\ast\hat{F}_{2})(u)=\int_{\mathcal{U}}\hat{F}_{1}(v)\alpha_{v}(\hat{F}%
_{2}(v^{-1}u))dv$, mapped by $\lambda^{\mathcal{M}}$ into $\lambda
^{\mathcal{M}}(\hat{F}_{1}\ast\hat{F}_{2})=\lambda^{\mathcal{M}}(\hat{F}%
_{1})\lambda^{\mathcal{M}}(\hat{F}_{2})$, or, as a von Neumann algebra
$\pi_{\alpha}(\mathcal{M})\vee(1\otimes\lambda(\mathcal{U}))$ generated by
$1\otimes\lambda(\mathcal{U})$ and by
\[
\pi_{\alpha}(\mathcal{M}):=\{\pi_{\alpha}(M):=Ad(U(W)^{\ast})(M\otimes
1);~M\in\mathcal{M}\}.
\]
These two versions are related by the mapping $\alpha(W):=Ad(U(W))$,
\[
\lambda^{\mathcal{M}}(L^{1}(\mathcal{U},\mathcal{M}))^{\prime\prime
}=(\mathcal{M}\otimes1)\vee\{U_{u}\otimes\lambda_{u};u\in\mathcal{U}%
\}\overset{\alpha(W)^{-1}}{\underset{\alpha(W)}{\rightleftarrows}}\pi_{\alpha
}(\mathcal{M})\vee(1\otimes\lambda(\mathcal{U})),
\]
which can be understood as the Schr\"{o}dinger and Heisenberg pictures: the
former $(\mathcal{M}\otimes1)\vee\{U_{u}\otimes\lambda_{u};u\in\mathcal{U}\}$
is in the Schr\"{o}dinger picture with \textit{unchanged }microscopic
observables $\mathcal{M}\otimes1$ and with the\textit{ coupling} $U_{u}%
\otimes\lambda_{u}$ \textit{to change macroscopic states}, while, in the
latter, all the coupling effects are concentrated in the observables
$\pi_{\alpha}(\mathcal{M})$ in contrast to the\textit{ kinematical changes} of
macroscopic\textit{ states} caused by $\lambda(\mathcal{U})$.

In the case of the instrument, the effects of the measurement coupling
$\widetilde{U}(V)$ are encoded in the form of \textit{macroscopic state}
\textit{changes} recorded in the spectrum of the non-trivial centre
$\mathfrak{Z}(\mathcal{M}\otimes\mathcal{A)}=\mathcal{A}=L^{\infty
}(Spec(\mathcal{A}))$ of $\mathcal{M}\otimes\mathcal{A}$, playing the same
roles as the order parameters to specify sectors in the inter-sectorial
context. For these reasons, the most natural physical essence of the formalism
in terms of an instrument $\mathcal{I}$ can be found in the
\textit{interaction picture}, whose\textit{ coupling} term $\widetilde
{U}(V)=(id\otimes\mathcal{F})U(W)^{\ast}(id\otimes\mathcal{F})^{-1}$ is
responsible for deforming the decoupled algebra $\mathcal{M}\otimes
\mathcal{A}$ into the above crossed product $\mathcal{M}\rtimes_{\alpha
}\mathcal{U}$.

To clarify the natural meaning of the above scheme, we note a useful analogy
of the duality coupling to the familiar \textit{complementarity of DNA
}between A(denine) and T(hymine) and between G(uanine) and C(ytosine),
repectively: the role of the coupling between $dE(\chi)$ and $\lambda_{\chi}$
in the K-T operator $\widetilde{U}(V)=\int_{\chi\in Spec(\mathcal{A})}%
dE(\chi)\otimes\lambda_{\chi}$ is just similar to that of the complementarity
of A-T and G-C, as the former implements the \textit{transcription} of the
data $\chi$ in the object system to the probe system in the form of
$\lambda_{\chi}:\iota\rightarrow\chi$ similarly to the latter case.

At this point, we note that the above standard description of measurement
processes in terms of an instrument implicitly presupposes that the
\textit{quantum}-theoretical processes, $\xi\rightarrow\xi_{\chi}$ and
$\iota\rightarrow\chi$, taking place at the\textit{ microscopic} contact point
of the observed and the probe systems can be directly interpreted as the
measured data $\chi$ identifiable with a position of the measuring pointer
visible at the \textit{macroscopic level}. There exist certain mathematical
and/or physical gaps between these two levels which need be filled up: to
adjust theoretical descriptions to the realistic experimental situations, we
need to discuss how these changes of probe systems dynamically propagate into{
macroscopic} motions of the measuring pointer. This is just the problem of the
\textit{amplification} processes to amplify the invisible quantum state
changes in the probe system into the macroscopic data registered in some
visible form of suitable order parameters. (Continuuing the above analogy to
the DNA, the aspect of amplification can naturally be compared with the
process of \textit{PCR}[= polymer chain reaction] to amplify the sequential
data of DNA.) In the next section, we formulate its general and abstract
essence in mathematical terms, by which the notion of the instrument need be
supplemented. In view of the inevitable noises in the actual experiment
situations, it is also necessary to show how the relevant information survives
to reach the macroscopically visible level, which requires the estimates of
the disturbance terms in the form of adiabaticity condition as will be done in
\S \ref{Example}.

\section{Unified Description of Measurements and Amplifications}

\label{Amplification}

We note here such a remarkable property inherent in the regular representation
of $\widehat{\mathcal{U}}$ as the mutual quasi-equivalence, $\lambda^{\otimes
m}\approx\lambda^{\otimes n}$ ($\forall m,n\in\mathbb{N}$), among its
arbitrary tensor powers $\lambda^{\otimes n}:=(\widehat{\mathcal{U}}\ni
\gamma\longmapsto\underset{n}{\underbrace{\lambda_{\gamma}\otimes\cdots
\otimes\lambda_{\gamma}}}\in U(L^{2}(\widehat{\mathcal{U}}))^{\otimes n})$, as
seen by the repeated use of the intertwining relation $V(\lambda_{\gamma
}\otimes1)=(\lambda_{\gamma}\otimes\lambda_{\gamma})V$ of the K-T operator
$V$:
\begin{align*}
&  V_{N,N+1}\cdots V_{23}V_{12}(\lambda_{\gamma}\otimes1^{\otimes N})\\
&  =V_{N,N+1}\cdots V_{23}V_{12}((\lambda_{\gamma}\otimes1)\otimes
\cdots\otimes1)\\
&  =V_{N,N+1}\cdots V_{23}((\lambda_{\gamma}\otimes\lambda_{\gamma}%
)\otimes1\otimes\cdots\otimes1)V_{12}\\
&  =V_{N,N+1}\cdots V_{23}(\lambda_{\gamma}^{\otimes2}\otimes1^{\otimes
(N-1)})V_{12}\\
&  =\cdots=V_{N,N+1}\cdots V_{n,n+1}(\lambda_{\gamma}^{\otimes n}%
\otimes1^{\otimes(N-n+1)})V_{n-1,n}\cdots V_{23}V_{12}\\
&  =\lambda_{\gamma}^{\otimes(N+1)}V_{N,N+1}\cdots V_{23}V_{12}.
\end{align*}
On this basis, we can formulate a dynamical process of amplification
\cite{Oji06} in terms of a unitary action $T_{N}$ of $\mathbb{N}$ on the
tensor algebra $\mathcal{M}\otimes(\otimes L^{\infty}(\widehat{\mathcal{U}}))$
with $\otimes L^{\infty}(\widehat{\mathcal{U}}):=\underset{N}{\underset
{\longrightarrow}{\lim}}\underset{N}{\underbrace{L^{\infty}(\widehat
{\mathcal{U}})\otimes L^{\infty}(\widehat{\mathcal{U}})\otimes\cdots\otimes
L^{\infty}(\widehat{\mathcal{U}})}\text{ }}$defined by%

\begin{align*}
&  T_{N}(A\otimes f_{2}\otimes\cdots\otimes f_{N+1})\\
&  :=\widetilde{U}(V)_{12}^{\ast}V_{23}^{\ast}\cdots V_{N,N+1}^{\ast}(A\otimes
f_{2}\otimes\cdots\otimes f_{N+1})V_{N,N+1}\cdots V_{23}\widetilde{U}%
(V)_{12}\\
&  =Ad(\widetilde{U}(V)_{12}^{\ast})Ad(V_{23}^{\ast})\cdots Ad(V_{N,N+1}%
^{\ast})(A\otimes f_{2}\otimes\cdots\otimes f_{N+1})\\
&  =Ad(\widetilde{U}(V)^{\ast})(A\otimes Ad(V^{\ast})(f_{2}\otimes Ad(V^{\ast
})(\cdots\otimes Ad(V^{\ast})(f_{N}\otimes f_{N+1})))\cdots)\\
&
\text{\ \ \ \ \ \ \ \ \ \ \ \ \ \ \ \ \ \ \ \ \ \ \ \ \ \ \ \ \ \ \ \ \ \ \ \ \ \ \ \ for
}A\in\mathcal{M}\text{\ and }f_{i}\in L^{\infty}(\widehat{\mathcal{U}}),
\end{align*}
which is similar to the formulation of quantum Markov chain due to Accardi
\cite{Acc74}. When $\widehat{\mathcal{U}}$ is discrete, this process can be
seen in a more clear-cut way in the Schr\"{o}dinger picture:%
\begin{align*}
U_{N}  &  :=V_{N,N+1}\cdots V_{23}\widetilde{U}(V)_{12};\\
&  U_{N}(\xi\otimes|\iota\rangle^{\otimes N})=\sum_{\gamma\in Spec(\mathcal{A}%
)}c_{\gamma}V_{N,N+1}\cdots V_{34}V_{23}(\xi_{\gamma}\otimes|\gamma
\rangle\otimes|\iota\rangle\otimes\cdots\otimes|\iota\rangle)\\
&  =\sum_{\gamma\in Spec(\mathcal{A})}c_{\gamma}V_{N,N+1}\cdots V_{34}%
(\xi_{\gamma}\otimes|\gamma\rangle\otimes|\gamma\rangle\otimes|\iota
\rangle\otimes\cdots\otimes|\iota\rangle)\\
&  =\cdots=\sum_{\gamma\in Spec(\mathcal{A})}c_{\gamma}\xi_{\gamma}%
\otimes\left[  |\gamma\rangle^{\otimes N}\right]  ,
\end{align*}
where $\xi=\sum_{\chi\in Spec(\mathcal{A})}c_{\chi}\xi_{\chi}$ is a generic
state $\xi\in L^{2}(\mathcal{M})$ of the observed system. According to the
general basic idea of `quantum-classical correspondence', a classical
macroscopic object is to be identified with a \textit{condensed state of
infinite number of quanta}, as well exemplified by the macroscopic
magnetization of Ising or Heisenberg ferromagnets described by the aligned
states $|\uparrow\rangle^{\otimes N}$ of `infinite number' $N\gg1$ of
microscopic spins. Likewise, the states $|\iota\rangle^{\otimes N}
:=\underset{N}{\underbrace{|\iota\rangle\otimes|\iota\rangle\otimes
\cdots\otimes|\iota\rangle}}$ and $|\gamma\rangle^{\otimes N}:=\underset
{N}{\underbrace{|\gamma\rangle\otimes|\gamma\rangle\otimes\cdots\otimes
|\gamma\rangle}}$ (with $N\gg1$) can physically be interpreted as representing
macroscopic positions of the measuring pointer corresponding, respectively, to
the initial and final probe states parametrized by $\iota$ and $\gamma$. Thus
the above repeated action $V_{N,N+1}\cdots V_{23}\widetilde{U}(V)_{12}$ of the
K-T operator $V$ describes a cascade process or a domino effect\ of
`decoherence',\ which, triggered by the initial data $\xi_{\gamma}$ of the
observed system, amplifies a probe state change $|\iota\rangle\rightarrow
|\gamma\rangle$ at the microscopic end of the apparatus to be transferred into
the macroscopic classical motion $\iota\rightarrow\gamma$ of the measuring pointer.

In view of the above aspects, we define a unified version of the instrument
combined with the amplification process:
\[
\widehat{\mathcal{I}}_{N}(\Delta|\omega_{\xi})=(\omega_{\xi}\otimes
m_{\mathcal{U}}{}^{\otimes N})(U_{N}^{\ast}((-)\otimes\chi_{\Delta}^{\otimes
N})U_{N}),
\]
in terms of which we can give an affirmative answer to the question posed at
the end of the previous section, \S 2, concerning the realistic meaning of the
quantity $\Delta$ as the actual data to be read out from the measuring
pointer. To this end, we show the equality
\[
\mathcal{I}(\Delta|\omega_{\xi})=\widehat{\mathcal{I}}_{N}(\Delta|\omega_{\xi
})
\]
between the usual and the above instruments as follows: assuming the
discreteness of $\widehat{\mathcal{U}}$ for simplicity, we calculate for
$B\in\mathcal{M}$,
\begin{align*}
&  \widehat{\mathcal{I}}_{N}(\Delta|\omega_{\xi})(B)=(\omega_{\xi}\otimes
m_{\mathcal{U}}{}^{\otimes N})(U_{N}^{\ast}(B\otimes\chi_{\Delta}^{\otimes
N})U_{N})\\
&  =\left(  \underset{\chi_{1}\in Spec(\mathcal{A})}{{\sum}}c_{\chi_{1}}%
^{\ast}\langle\xi_{\chi_{1}}|\otimes\langle\chi_{1}|^{\otimes N}\right)
\left(  B\otimes\chi_{\Delta}^{\otimes N}\right)  \left(  \underset{\chi
_{2}\in Spec(\mathcal{A})}{{\sum}}c_{\chi_{2}}|\xi_{\chi_{2}}\rangle
\otimes|\chi_{2}\rangle^{\otimes N}\right) \\
&  ={\sum}_{\chi\in\Delta}|c_{\chi}|^{2}\langle\xi_{\chi}|B|\xi_{\chi}%
\rangle\chi_{\Delta}(\chi)^{N}={\sum}_{\chi\in\Delta}|c_{\chi}|^{2}\langle
\xi_{\chi}|B|\xi_{\chi}\rangle\\
&  =\mathcal{I}(\Delta|\omega_{\xi})(B)
\end{align*}
which reduces for $\Delta=\{\gamma\}(\subset Spec(\mathcal{A}))$ to such a
familiar result as
\begin{align}
\mathcal{I}(\{\gamma\}|\omega_{\xi})(B)  &  =\widehat{\mathcal{I}}%
_{N}(\{\gamma\}|\omega_{\xi})(B)=|c_{\gamma}|^{2}\langle\xi_{\gamma}%
|B|\xi_{\gamma}\rangle;\nonumber\\
p(\{\gamma\}|\omega_{\xi})  &  =\mathcal{I}(\{\gamma\}|\omega_{\xi
})(\mathbf{1})=\widehat{\mathcal{I}}_{N}(\{\gamma\}|\omega_{\xi}%
)(\mathbf{1})\nonumber\\
&  =|c_{\gamma}|^{2}\text{\ \ for}\ \forall N\in\mathbb{N},\gamma\in
Spec(\mathcal{A}).\text{ } \label{LDP}%
\end{align}
Since $|c_{\gamma}|^{2}$ gives precisely the probability of finding a
\textit{macroscopic} state $|\gamma\rangle^{\otimes N}$, we have observed just
the agreement of the probability distributions between the one arising from
the microscopic system-probe coupling and the final result realized through
the amplification process. This fact ensures the pertinence of instruments for
the description of measurements, giving a clear-cut version of
quantum-classical correspondence.

\subsection*{Infinite divisibility and L\'{e}vy process}

The unitarity\ of the above amplification process is guaranteed by the
quasi-equivalence relations among arbitrary tensor powers $\lambda^{\otimes
n}=\lambda\otimes\cdots\otimes\lambda$ of the regular representation $\lambda$
of $\widehat{\mathcal{U}}$. It can also explain the possibility of the
recurrent quantum interference even after the contact of a quantum system with
the measuring apparatus when the number $N$ of repetition need not be regarded
as a real infinity. This point is evident from Eq.(\ref{LDP}) which is valid
independently of $N\in\mathbb{N}$. In general, the problem as to whether the
situation is made `completely'\ classical or not depends highly on the
relative configurations among many large or small numbers, which can
consistently be described in the framework of the non-standard analysis (see,
for instance, \cite{OjiOza}).

In close relation to this, it is also interesting to note that the above
amplification process is related to a L\'{e}vy process through its `infinite
divisibility'\ as follows: similarly to the affine property $f(\lambda x+\mu
y)=\lambda f(x)+\mu f(y)$ ($\forall\lambda,\mu>0$) of a map $f$ defined on a
convex set following from the additivity $f(x+y)=f(x)+f(y)$, we can
extrapolate the relation $\lambda^{m}\thickapprox\lambda^{n}$ ($\forall
m,n\in\mathbb{N}$) into $\lambda\thickapprox\lambda^{n/m}$, which means the
infinite divisibility\ $(Ad(V))^{t+s}\thickapprox(Ad(V))^{t}(Ad(V))^{s}$
($t,s>0$) of the process induced by the above transformation. In this way, we
see that simple individual measurements with definite measured values are
connected without gaps with discrete and/or continuous repetitions of
measurements \cite{OjiTan92}. If this formulation exhausts the essence of the
problem, the remaining tasks reduce to its physical and/or technical
implementation through suitable choices of the media connecting the
microscopic contact point between the system and the apparatus to the
measuring pointer. In such contexts, we need to examine some aspects
concerning the stability of the information transmitted from microscopic to
macroscopic levels, as will be seen in the next section.

\section{Example Case of Stern-Gerlach Experiment}

\label{Example}

In this section we apply the scheme developed so far to the experimental
situation of Stern-Gerlach type to check the validity of its general essence
and to attain a deeper understanding of it through the concrete example. We
will find also the necessity of some generalization or modification for
adapting the scheme to actual situations. The essence of Stern-Gerlach
experiments\footnote{Suggested by O.Stern \& W.Gerlach in 1922}
\cite{Bohm,Tomonaga} can be found in the coupling between the (spin and/or
orbital) angular momentum of the quantum particles (such as atoms or
electrons) and the inhomogeneous external magnetic field, according to which
the \textit{microscopic} differences in the\textit{ }quantized directions of
angular momentum are amplified into the \textit{macroscopic} distance of the
arriving points of the particle. For simplicity, we consider here the spin
{\boldmath$\sigma$} $=(\sigma_{x},\sigma_{y},\sigma_{z})$ of an electron (with
spin $s=1/2$), whose associated magnetic moment $\mu${\boldmath$\sigma$
}couples to the magnetic field via the interaction term $\mu${\boldmath$\sigma
$}$\cdot\mathbf{B}(\mathbf{x})$: through the $\mathbf{x}$-dependence of
$\mathbf{B}(\mathbf{x})$ due to its inhomogeneity, this coupling causes the
orbital change of the electron according to its spin direction (\textit{up} or
\textit{down}) with respect to the defined axis (see
Fig.\newblock\ref{setting} in \S \ref{S-G}). Thus the magnetic field
$\mathbf{B}(\mathbf{x})$ is seen to play a double role; the coupling $\mu
${\boldmath$\sigma$}$\cdot\mathbf{B}$ causes, on the one hand, the
\textit{spectral decomposition} of the quantum spin {\boldmath$\sigma$}, and
it causes, on the other hand, the \textit{amplification process} through its
dependence on $\mathbf{x}$. Through the process, we can `see' the quantum spin
variable of the electron as the separation of its spatial orbits (or, more
directly, the arriving points on the screen). Thus the two states
$|\!\uparrow\rangle$ and $|\!\downarrow\rangle$, respectively, of spin up and
down, can be distinguished through the amplification process caused by the
Stern-Gerlach measurement apparatus.

In \S \ref{Amplification} the amplification process was formulated in its
idealized abstract form in terms of the homogeneous repetition by a K-T
operator. In the present case of Stern-Gerlach experiment, however, the
coupling between the electron and the inhomogeneous magnetic field depends on
the position of the moving electron, owing to which the unitary coupling term
$V=\exp[\frac{it}{\hbar}\mu${\boldmath$\sigma$}$\otimes\mathbf{B}%
(\mathbf{x})]$ depends on the position $\mathbf{x}$ of the electron along its
trajectory. At the same time, any amplification processes cannot get rid of
\textit{noise effects} to disturb the ideal separations between upward and
downward electron beams corresponding to macroscopically distinguishable
states $|\!\uparrow\rangle^{\otimes\infty}$ and $|\!\downarrow\rangle
^{\otimes\infty}$, respectively. For these reasons, it is necessary to examine
whether the possible spin-flips during the travel of electron through the
magnetic field can sufficiently be suppressed. Otherwise, frequent spin-flips
may destroy the meaningful connection between the spin variables of the
electrons and the points on the screen to detect them. Therefore, to ensure
the distinguishability and the stability in the separations of final results,
some physical conditions need be supplemented to ensure that these `error
probability' is small enough. This can be understood as a kind of
`adiabaticity condition' related with the validity of adiabatic approximation
to treat the varying and fluctuating background field.

\subsection{Applying the scheme to Stern-Gerlach experiment}

\label{S-G}

\begin{figure}[th]
\begin{center}
\includegraphics{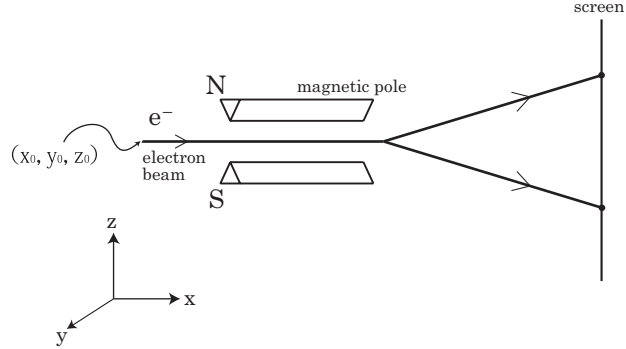}
\end{center}
\caption{{\protect\small The settings of Stern-Gerlach experiment with single
pair of magnetic poles}}%
\label{setting}%
\end{figure}

The standard setting of the Stern-Gerlach experiment is shown below (see
Fig.\newblock\ref{setting} to illustrate the apparatus); we prepare a given
type of metal which emits the electron beam through the thermal oscillation.
The thermal electronic beam enters in the inhomogeneous magnetic field
$\mathbf{B}(\mathbf{x})$ generated between magnetic poles which covers a
spatial region with a length scale of the order of a meter. The orbital motion
of each electron is bent upward or downward according to the directions of its
spin coupled to the magnetic field; the microscopic state determined by the
direction of electron spin as an invisible \textit{internal} degree of freedom
is thus converted into the visible macroscopic form of spatial separations of
the spots on the screen caused by the electrons.

From here, we focus on the situation for detecting the spin direction
consisting of an electron with spin $s=1/2$, mass $m$, charge $e$, magnetic
moment $\mu${(}$=e\hbar/2{\mu}_{0}mc$ with magnetic permeability ${\mu}_{0}$
of vacuum{) }and of the external magnetic field whose direction is supposed to
be fixed in the $z$-axis. 

For applying our general scheme, we should proceed in the following steps:

\begin{description}
\item 0) To find the algebra which describes the physical system.

\item 1) To extract the basic ingredients relevant to Micro-Macro duality
(MASA, unitary group and their duals) from the algebra found in 0).

\item 2) To identify the K-T operator in terms of these ingredients.
\end{description}

\paragraph{0) To find the algebra which describes the physical system.}

The physical variables of the electron constitute the algebra $M_{2}%
(\mathbb{C})\otimes B(L^{2}(\mathbb{R}^{3}))$ consisting of the spin variables
$M_{2}(\mathbb{C})=Lin(\sigma_{x},\sigma_{y},\sigma_{z})^{\prime\prime}$ and
the algebra $B(L^{2}(\mathbb{R}^{3}))$ of the canonical commutation relations
(a CCR algebra, for short, or, a Heisenberg algebra) generated, respectively,
by Pauli matrices $\sigma_{i}$ and by the spatial coordinates $x,y,z$ and the
momenta $p_{x},p_{y},p_{z}$. According to the general framework in
\S \ref{Intro}, we can take $\mathcal{M}=M_{2}(\mathbb{C})\otimes
B(L^{2}(\mathbb{R}^{3}))$ as the algebra describing the system to be observed
(as a von Neumann algebra of type I).

\paragraph{1) To extract basic ingredients relevant to Micro-Macro duality.}

We can find the MASA as%
\[
\mathcal{A}=\mathcal{A}^{\prime}=Diag(2,\mathbb{C})\otimes L^{\infty
}(\mathbb{R}^{3}),
\]
up to unitary conjugacy, where $Diag(2,\mathbb{C})$ denotes the set of
$2\times2$ diagonal matrices $\left(
\begin{array}
[c]{cc}%
\alpha & 0\\
0 & \beta
\end{array}
\right)  $ ($\alpha,\beta\in\mathbb{C}$). This algebra is generated by the
group $\mathcal{U}=\mathcal{U}(\mathcal{A})$ of its unitary elements:%
\[
\mathcal{U}=\mathcal{U}(\mathcal{A})=\mathbb{T}^{2}\otimes\mathcal{U}%
(L^{\infty}(\mathbb{R}^{3}))=\mathbb{T}^{2}\otimes L^{\infty}(\mathbb{R}%
^{3},\mathbb{T}).
\]
The dual objects are also determined as follows: \newline\newline\qquad
Spectrum:$~Spec(\mathcal{A})=\{\pm1\}\times\mathbb{R}^{3};$ \newline%
\newline\qquad Dual group:~$\widehat{\mathcal{U}}=\mathbb{Z}^{2}\otimes
L^{1}(\mathbb{R}^{3},\mathbb{Z}),$\newline\newline where $L^{1}(\mathbb{R}%
^{3},\mathbb{Z})$ consists of compactly supported $\mathbb{Z}$-valued step
functions on $\mathbb{R}^{3}$, namely, each element $f\in L^{1}(\mathbb{R}%
^{3},\mathbb{Z})$ takes a constant integer value $c_{i}\in\mathbb{Z}$ on each
$\Delta_{i}$ of a finite number of non-intersecting Borel sets $\Delta
_{1},\cdots,\Delta_{r}$ in $\mathbb{R}^{3}$ and vanishes outside of
$\displaystyle{\bigcup} _{i=1,\cdots,r}\Delta_{i}$: $f(x)=\left\{
\begin{array}
[c]{c}%
c_{i}\text{ for }x\in\Delta_{i},\\
0\text{ otherwise. }%
\end{array}
\right.  $

We note that it is possible to extract the information on the spin degrees of
freedom of the observed system from the spin algebra only, ignoring the
orbital part described by the CCR. In this context, the relevant MASA
$\{\sigma_{z}\}^{\prime\prime}$ is just the \textit{Cartan subalgebra} of the
Lie algebra $\mathfrak{su}(2)=Lin(\sigma_{x},\sigma_{y},\sigma_{z})$ (as is
familiar in the theory of semi-simple Lie algebras), where the spectrum
$\mathbb{Z}_{2}=\{\pm1\}\subset Spec(\mathcal{A})$ can be identified with its
root system. Physically they correspond to the spin up/down states with
respect to the $z$-axis. In contrast to $Spec(\mathcal{A})$ having no identity
element in itself, we can identify the unit element $(0,0)\in\mathbb{Z}^{2}$
of the dual group $\mathbb{Z}^{2}$ as the \textit{neutral position} of the
measuring system, which can also be identified with the Haar measure
$dt_{1}dt_{2}$ of $\mathbb{T}^{2}$ or the constant function $1$ on
$\mathbb{T}^{2}$. While this neutral position does not exist as a position of
measuring pointer, operationally it represents a situation of \textit{no click
}on either of upper or lower detector. Generic states of electron spin to be
measured are represented by arbitrary superpositions $c_{+}|\!\uparrow
\rangle+c_{-}|\!\downarrow\rangle$ $(c_{+},c_{-}\in\mathbb{C},\left\vert
c_{+}\right\vert ^{2}+\left\vert c_{-}\right\vert ^{2}=1)$ of two eigenstates
$|\!\uparrow\rangle,|\!\downarrow\rangle$ of $\sigma_{z}$. According to the
result in \S \ref{Measurement}, the coefficient $c_{\pm}$ gives the transition
amplitude from the above `state'\ of neutral position (of the measuring
pointer) to either of the `amplified'\ macroscopic states $|\!\uparrow
\rangle^{\otimes N}$ and $|\!\downarrow\rangle^{\otimes N}$.

\paragraph{2) To identify the K-T operator.}

Our aim here is to understand the role of the coupling Hamiltonian $\mu
${\boldmath$\sigma$}$\otimes\mathbf{B}(\mathbf{x})$ in relation with a K-T
operator and its associated instrument. For this purpose, we consider a
(trivial) vector bundle $E:=\mathbb{R}^{3}\times\mathbb{C}^{2}%
\twoheadrightarrow\mathbb{R}^{3}$ over a base space $\mathbb{R}^{3}$ spanned
by the electron coordinates $\mathbf{x}$ with a fibre $\mathbb{C}^{2}\ni
\psi(\mathbf{x})$ describing spin states of the electron at $\mathbf{x}$; $E$
has group actions on its base space and its standard fibre, respectively, by
the 3-dimensional motion group $M(3)=\mathbb{R}^{3}\underset{Ad}{\rtimes
}SU(2)=:G$ and by the spin rotations $SU(2)=:H$, where $\underset{Ad}{\rtimes
}$ means the semi-direct product w.r.t. the adjoint action of $SU(2)$ on
$\mathbb{R}^{3}\simeq\{X\in M_{2}(\mathbb{C});Tr(X)=0,X^{\ast}=X\}$. It is
important here to note that $E$ is a homogeneous bundle over the homogeneous
space $G/H\simeq\mathbb{R}^{3}$, according to which a representation of $G$
can be induced from that of its subgroup $H$. Therefore, the geometry involved
in the Stern-Gerlach experiment (as an intra-sectorial version) can be related
to the measurement scheme \cite{Unif03, QBIC07} for a \textit{sector bundle}
$G\underset{H}{\times}\hat{H}\twoheadrightarrow G/H$ over $G/H$ consisting of
the \textit{degenerate vacua} associated to a spontaneous symmetry breaking of
$G$ into an unbroken subgroup $H$ with the standard fibre $\hat{H}$ describing
the sector structure associated with $H$:%
\[
\left[
\begin{array}
[c]{ccc}%
\mathcal{M}^{H}\rtimes\widehat{G}\simeq\mathcal{M}\rtimes\widehat{(H\backslash
G)} & \Longrightarrow &
\begin{array}
[c]{c}%
\text{read-out data in }Spec(\text{centre})\\
\text{(I)}=G/H\text{: degenerate vacua}%
\end{array}
\\
\Uparrow\text{\ \ \ \ \ \ \ \ \ \ \ \ \ \ \ \ \ \ } &  & \\%
\begin{array}
[c]{c}%
\hat{G}\curvearrowright\lbrack\mathcal{M}\rtimes H\simeq\mathcal{M}^{H}]\\
\text{: coupling (I) \ \ \ \ \ \ \ \ }%
\end{array}
& \Longrightarrow &
\begin{array}
[c]{c}%
\text{read-out data in }Spec(\text{centre})\\
\text{(II)}=\hat{H}\text{: sectors on a vacuum}%
\end{array}
\\
\text{ \ \ \ \ \ \ \ \ \ \ \ \ \ \ }\Uparrow &  & \\
\text{coupling (II): }H\curvearrowright\mathcal{M} &  &
\end{array}
\right]  .
\]
The interpretation of each step of (I) and (II) in this diagram is just in
parallel with our measurement scheme: the unbroken subgroup $H$ acts on the
algebra $\mathcal{M}$ of observables of the system through the coupling (II),
according to which the associated sector structure over a fixed vacuum can be
read off (II) in terms of $\hat{H}$ realized as the spectrum of the centre of
$\mathcal{M}\rtimes H$, and, similarly, the coupling (I) to implement the
co-action of $G$ on the crossed product $\mathcal{M}\rtimes H$ makes it
possible to observe the sector structure (I) of the degenerate vacua
parametrized by $G/H$. From this viewpoint, the interaction Hamiltonian
{\boldmath$\sigma$}$\otimes\mu\mathbf{B}(\mathbf{x})=${$\sigma_{z}$}%
$\otimes\mu B_{z}(\mathbf{x})=\left(
\begin{array}
[c]{cc}%
\mu B_{z}(\mathbf{x}) & 0\\
0 & -\mu B_{z}(\mathbf{x})
\end{array}
\right)  $ can be interpreted as follows: the coupling term $\exp[\frac
{it}{\hbar}${$\sigma_{z}$}$\otimes\mu B_{z}]$ exhibits, via spectral
decomposition, the `sector' structure $\sigma_{z}=\pm1$ parametrized by the
roots $\pm1$ of $H=SU(2)$ similarly to the above (II) within a fibre. When we
recall the $\mathbf{x}$-dependence of $B_{z}=B_{z}(\mathbf{x}),\mathbf{x}%
\in\mathbb{R}^{3}=G/H$, the aspects (I) of the degenerate vacua as condensed
states shows up in relation with the base space $G/H=\mathbb{R}^{3}$. To see
this, we consider such an approximation of the inhomogeneus magnetic field
$B_{z}(\mathbf{x})$ as
\[
B_{z}(\mathbf{x})\simeq B_{0}+\dfrac{\partial B_{z}}{\partial z}z=B_{0}%
+B_{1}z.
\]
This allows us to interpret the above coupling term $\exp[\frac{i}{\hbar
}\Delta t${$\sigma_{z}$}$\otimes\mu B_{z}]$ (for a time interval $\Delta t$)
as another K-T operator relevant to (I):
\begin{align*}
\exp[\frac{i}{\hbar}\Delta t{\sigma_{z}}\otimes\mu B_{z}(\mathbf{x})]  &
=\left(
\begin{array}
[c]{cc}%
\exp[\frac{i}{\hbar}\mu B_{z}(\mathbf{x})\Delta t] & 0\\
0 & \exp[-\frac{i}{\hbar}\mu B_{z}(\mathbf{x})\Delta t]
\end{array}
\right) \\
&  \simeq e^{\frac{i}{\hbar}\sigma_{z}\mu B_{0}\Delta t}\left(
\begin{array}
[c]{cc}%
e^{\frac{i}{\hbar}\mu B_{1}z\Delta t} & 0\\
0 & e^{-\frac{i}{\hbar}\mu B_{1}z\Delta t}%
\end{array}
\right)  ,
\end{align*}
which describes the (co-)action of the $z$-axis $\mathbb{R}\subset G/H$ on
$\mathcal{M}$ to generate $\mathcal{M}\rtimes\widehat{(H\backslash G)}$ (an
\textit{augmented algebra }introduced in \cite{Unif03}). To understand this,
it is sufficient to note that the exponent $\pm\frac{i}{\hbar}\mu B_{1}z\Delta
t$ of matrix elements $\exp(\pm\frac{i}{\hbar}\mu B_{1}z\Delta t)$ in the
above coupling unitary $\exp(\frac{i}{\hbar}\Delta t{\sigma_{z}}\otimes\mu
B_{z})$ can be seen as the spectral value of the K-T operator $\exp(\frac
{i}{\hbar}\hat{p}_{z}\otimes\hat{z})=\int dE(p_{z})\otimes\hat{\lambda}%
_{p_{z}}$ corresponding to the (generalized) eigenvalue $p_{z}=\pm\mu
B_{1}\Delta t$ of the momentum operator $\hat{p}_{z}$:
\begin{equation}
\Delta p_{z}=\pm\mu B_{1}\Delta t\text{ \ \ \ : corresponding to the
eigenvalue }\pm1\text{ of }\sigma_{z}, \label{eqn:MAIN2}%
\end{equation}
In the context of group representations, two representations of $G=\mathbb{R}%
^{3}\underset{Ad}{\rtimes}SU(2)$ are induced from the two representations of
$H=SU(2)$ corresponding to the eigenvalues $\sigma_{z}=\pm1$, which are
restricted to another subgroup $\mathbb{R}^{3}$ and then to the $z$-axis
$\subset\mathbb{R}^{3}$, corresponding to (approximately) plane waves with
$p_{z}=\pm\mu B_{1}\Delta t$, which reach the upper/lower detectors,
respectively:
\[%
\begin{array}
[c]{ccc}
& \text{\ \ \ \ }G=\mathbb{R}^{3}\underset{Ad}{\rtimes}SU(2) & \\
&  & \\
& \text{\ \ \ \ \ }\text{(induction : ) }\nearrow
\text{\ \ \ \ \ \ \ \ \ \ \ \ }\searrow\text{ (: restriction)} & \\
&  & \\
& \sigma_{z}\longrightarrow\text{\ \ \ }G/\mathbb{R}^{3}=SU(2)\underset
{\text{Helgason duality}}{\longleftrightarrow}G/H=\mathbb{R}^{3}%
\text{\ \ }\longrightarrow\hat{p}_{z}\text{\ .} &
\end{array}
\]
In this way, the spin {\boldmath$\sigma$} and the orbital motion described by
$\mathbf{x},\mathbf{p}$ are coupled by the inhomogeneity of the external
magnetic field $B_{z}(\mathbf{x})\simeq B_{0}+B_{1}z$, according to which the
microscopic directions $\sigma_{z}=\pm1$ of the former is \textit{amplified}
into the macroscopic directions $p_{z}=\pm\mu B_{1}\Delta t$ in the orbital
motion. These latter directions can be understood as the `amplified'\ states,
$|\!\uparrow\rangle^{\otimes N}$ and $|\!\downarrow\rangle^{\otimes N}$ with
the upper/lower points on the target screen.

\begin{comment}
The `Helgason duality' above is a special case of the duality between the two
homogeneous spaces, $K\backslash G$ and $G/H$, constituting a double fibration
$K\backslash G\twoheadleftarrow G\twoheadrightarrow G/H$, which plays
important roles in the context of Radon transforms \cite{Helg99}.
\end{comment}

It is remarkable that the coupling unitary $\exp\left[  \frac{it}{\hbar
}{\sigma_{z}}\otimes\mu B_{z}(\mathbf{x})\right]  $ characteristic of the
Stern-Gerlach experiment contains the two kinds of K-T operators, the one,
$\exp(i\dfrac{{\sigma_{z}}}{2}\otimes\hat{\theta})$, to couple the quantum
observable {\boldmath$\sigma$ with the angle variable }$\hat{\theta}=2t\mu
B_{z}/\hbar$ and the other one, $\exp(\frac{i}{\hbar}\hat{p}_{z}\otimes\hat
{z})$, corresponding to the translations $z\rightarrow z+a$ of $z$ due to the
$z$-dependence of $B_{z}(\mathbf{x})\simeq B_{0}+B_{1}z$, the latter of which
is responsible for the direct amplification of the former coupling. This
explains a dynamical mechanism to transcribe the information on the spin
direction into the momentum change in the orbital motion of the electron,
which allows us to achieve the quantitative estimation as shown above.

Aside from the Stern-Gerlach case, a unitary coupling of the similar nature
has been found in \cite{Emch}\footnote{The paper by Prof. G. Emch has been
brought to our attention by Prof. Ohya, to whom we express our gratitudes.}.
Our focus here is, however, to clarify the universal essence of such couplings
\textit{via external fields}, which seems impossible without the use of K-T operators.

\subsection{Adiabatic perturbation and adiabaticity condition}

\label{Ad}

In the above discussion for deriving the momentum change of the electron, we
neglected such secondary effects as the terms come from $B_{x}$ or
$\dfrac{\partial B_{x}}{\partial z}$. Since these effects are outside the
scope of the above \textit{ideal} situation of amplification, we need to
estimate them as correction terms in the next step. Without the necessity to
develop the general method for treating these secondary terms, we already know
some of typical methodology for these estimation; in some cases (including the
Stern-Gerlach case) it would be called `adiabaticity conditions'. For
Stern-Gerlach experiment, this condition can be interpreted as the one under
which the effect of \textit{spin-flips} caused by the factor $\dfrac{\partial
B_{x}}{\partial z}$ remains small enough compared with that of $B_{z}$. In
this section, we confirm that the adiabaticity condition surely gives the
consistency in the present context by an elementary discussion.

`Adiabatic perturbation'\ originally means a coupling of a quantum system with
an external force which changes the system slowly enough in comparison to the
typical time scales of intrinsic transitions among quantum states but whose
changes \textit{along the direction of condensed order parameters} can
eventually accumulate into a visible size. The general essence of the
adiabaticity can be formulated in such a condition as $|U_{fi}|\ll1$, in terms
of the rate of change of the matrix elements of Hamiltonian $H$ defined by
\[
U_{fi}:=\frac{\hbar}{(E_{f}-E_{i})^{2}}\left(  \frac{\partial H}{\partial
t}\right)  _{fi},
\]
between the initial and final states with the energies $E_{i}$ and $E_{f}$,
respectively. The physical meaning of the quantity $U_{fi}$ can be understood
by the following reformulation of it:
\[
U_{fi}=\dfrac{(\Delta H)_{fi}}{(\Delta E)_{fi}}=\frac{\left(  \dfrac{\partial
H}{\partial t}\Delta t\right)  _{fi}}{E_{f}-E_{i}}=\frac{\left(
\dfrac{\partial H}{\partial t}\right)  _{fi}{\omega}_{fi}^{-1}}{E_{f}-E_{i}},
\]
with $\Delta t:=\dfrac{\hbar}{E_{f}-E_{i}}=:{\omega}_{fi}^{-1}$ which sets up
the standard time scale for the comparison. The requirement $|U_{fi}|\ll1$ can
now be understood as the self-consistency condition for a process to change
the values of the order parameters describing a given inter-sectorial
structure of the quantum-classical composite system, without destroying the
whole sector structure: if the change rate $\dfrac{\partial H}{\partial t}$ of
the Hamiltonian is very small, it should be almost perpendicular to the main
`tangential direction'\ of the changes caused by the external force in favour
of the change in the order parameters. Therefore, $\Delta t$ can be
interpreted as the `almost intrinsic'\ time scale of the microscopic motions
of the intra-sectorial quantum system put in a background with slowly changing
order parameters, in which $\omega_{fi}$ can represent, for instance, the
frequency of the light emitted in the transition. Then the numerator in
$U_{fi}=\dfrac{(\Delta H)_{fi}}{(\Delta E)_{fi}}$ is the change $(\Delta
H)_{fi}=\left(  \dfrac{\partial H}{\partial t}\Delta t\right)  _{fi}$ of the
matrix element of $H$ from the initial $i$ to final states $f$ \textit{caused
by the adiabatic perturbation} during the time interval $\Delta t$, which is
to be compared with the denominator $(\Delta E)_{fi}=E_{f}-E_{i}$ given by the
energy\textit{\ }difference \textit{almost intrinsic to the quantum system}.

Going back to the Stern-Gerlach case, the interaction Hamiltonian is given by
\begin{equation}
\widetilde{H_{I}}=\frac{e\hbar}{2\mu_{0}mc}\mathbf{B}{\cdot}%
\text{\boldmath$\sigma$}=\frac{e\hbar}{2\mu_{0}mc}(B_{x}{\sigma}_{x}%
+B_{z}{\sigma}_{z}),\nonumber
\end{equation}
The decomposition of the external magnetic field into its $z$-component
$B_{z}$ and the remaining $B_{x}$ can be understood as the one into the
directions to preserve and to disturb the sector structure according to the
eigenvalues of $\sigma_{z}$. Therefore, the dominant term in this Hamiltonian
to disturb the spin direction due to the spin-flips is identified with
\[
H_{I}=\frac{e\hbar}{2\mu_{0}mc}B_{x}{\sigma}_{x}.
\]
The size of the effect due to this term should be estimated to preserve the
visibility aspect due to $B_{z}$.

As each trajectory of electron can be considered as a smooth curve in
$\mathbb{R}^{3}$ parameterized by the \textit{time} parameter $t$, the time
derivative of $H_{I}$ is calculated as
\[
\frac{\partial H_{I}}{\partial t}=\frac{e\hbar}{2\mu_{0}mc}\frac{dx}{dt}%
\frac{\partial B_{x}}{\partial x}{\sigma}_{x}.
\]
Here we introduce an approximation $B_{x}\simeq B_{x}(z=0)+\dfrac{\partial
B_{x}}{\partial z}z$. In terms of a basis of eigenstates of $\sigma_{z}$, we
can estimate and obtain a representation of off-diagonal matrix elements%
\[
\left(  \frac{\partial H_{I}}{\partial t}\right)  _{fi}=\frac{e\hbar}{2\mu
_{0}mc}vz\frac{{\partial}^{2}B_{x}}{\partial x \partial z}\int\overline{{\psi
}_{f}(x)}{\sigma}_{x}{\psi}_{i}(x)dx,
\]
under the assumption that the velocity $dx/dt$ of the electron can be replaced
by the typical velocity $v$ of thermal electrons. Owing to the first condition
for $\partial H/\partial t$ to be adiabatic, the derivative of the external
magnetic field can be approximated in the context of the estimate by
${\partial B_{x}}/{\partial z}\sim B_{z}/(\Delta x)$, where $\Delta x$
represents the range in which the magnetic field exists.

In terms of the Larmor frequency of the thermal electron $\omega=eB_{z}%
/2\mu_{0}mc$, the changing rate in which we are interested is essentially
given by
\begin{align}
U_{fi}  &  =\frac{vz{\frac{\partial}{\partial x}}\left(  \frac{B_{z}}{\Delta
x}\right)  }{\omega B_{z}}\nonumber\\
&  =v\frac{z}{\omega\Delta x}\frac{1}{B_{z}}\frac{\partial B_{x}}{\partial
z}.\nonumber
\end{align}
in the use of the rotation-free condition $\dfrac{\partial B_{z}}{\partial
x}=\dfrac{\partial B_{x}}{\partial z}$ of the magnetic field $\mathbf{B}$.
Thus the adiabaticity condition $|U_{fi}|\ll1$ can be written down as
\begin{equation}
\frac{\partial B_{x}}{\partial z}\ll\frac{{\omega}}{v}B_{z}.
\end{equation}
This inequality is nothing but the condition imposed on the arrangement of
external magnetic field in order to guarantee the ideal amplification of spin variables.

\section{Summary}

\label{Sum}

In this paper we have formulated a unified scheme of measurement and
amplification processes based on the notion of Micro-Macro duality. In this
context, the duality relation (or, in more general contexts, adjunction)
between $\mathcal{M}$ as the microscopic system and $Spec(\mathcal{A})$ as the
macroscopic observational data controlled by the K-T operator has played the
essential role, on the basis of which we have obtained a clear understanding
of how microscopic states are amplified into macroscopic level as discussed in
\S \ref{Amplification}. We hope that this essence of amplification processes
will shed some new lights on various problems involving different scales or
levels (especially, `Micro' and `Macro') such as the coexistence of different
phases and their boundaries, the problem of emergence of macroscopic
structures from microscopic worlds, and so on.

\section*{Acknowledgments}

One of the authors (I.~O.) would like to express his sincere thanks to
Prof.~M.~Ohya, Prof.~L.~Accardi and Prof.~T.~Hida for their encouragements.
Both of the authors are very grateful to Mr.~H.~Ando, Mr.~T.~Hasebe and
Mr.~H.~Saigo for their valuable discussions in the early stage of the work.

\end{document}